**BREWING DISCONTENT:** ***How U.S. Reciprocal Tariffs on Coffee Could Echo the Boston Tea Party***

MUHAMMAD SUKRI BIN RAMLI
Asia School of Business
Kuala Lumpur, Malaysia
Email: m.binramli@sloan.mit.edu

**Abstract**

This research employs quantitative techniques interpreted through relevant economic theories to analyze a proposed U.S. "Discounted Reciprocal Tariff" structure. Statistical modeling (linear regression) quantifies the policy's consistent 'discounted reciprocity' pattern, which is interpreted using a Game Theory perspective on strategic interaction. Machine learning (K-Means clustering) identifies distinct country typologies based on tariff exposure and Economic Complexity Index (ECI), linking the policy to Economic Complexity theory. The study's primary application focuses on the major coffee exporting sector, utilizing simulation modeling grounded in principles of demand elasticity and substitution to project potential trade flow impacts. Specifically, for coffee, this simulation demonstrates how the proposed tariff differentials can induce significant substitution effects, projecting a potential shift in U.S. import demand away from high-tariff origins toward lower-tariff competitors. This disruption, stemming from the tariffs impacting exporting countries, is projected to ultimately increase coffee prices for consumers in the United States. Findings throughout are contextualized within Political Economy considerations. Overall, the study demonstrates how integrating regression, clustering, and simulation with economic theory—exemplified through the coffee sector analysis—provides a robust framework for assessing the potential systemic impacts, including consumer price effects, of strategic trade policies.

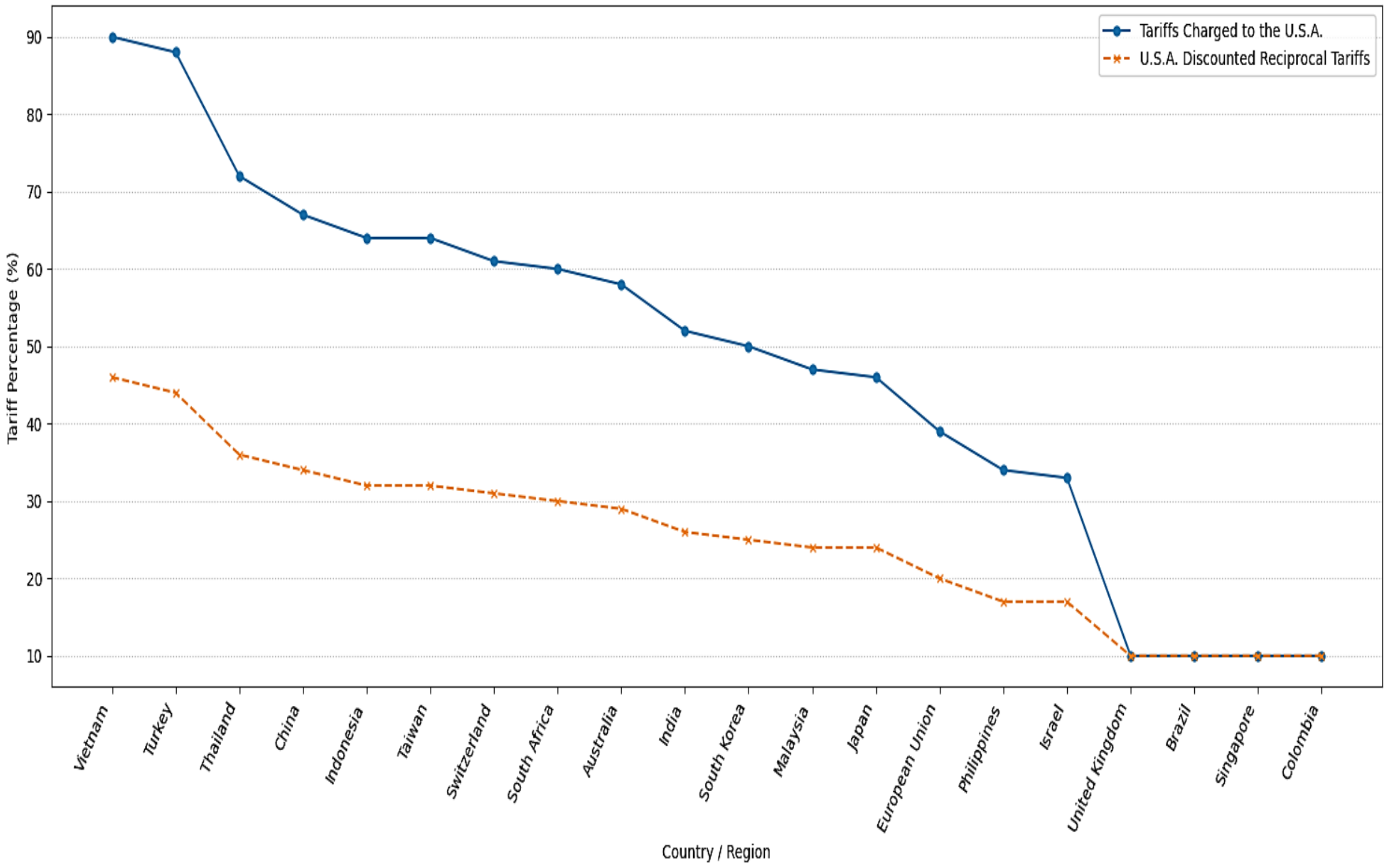

Figure 1: 2025 Tariffs Charged to the USA vs. USA Discounted Reciprocal Tariffs (Top Trading Partners)

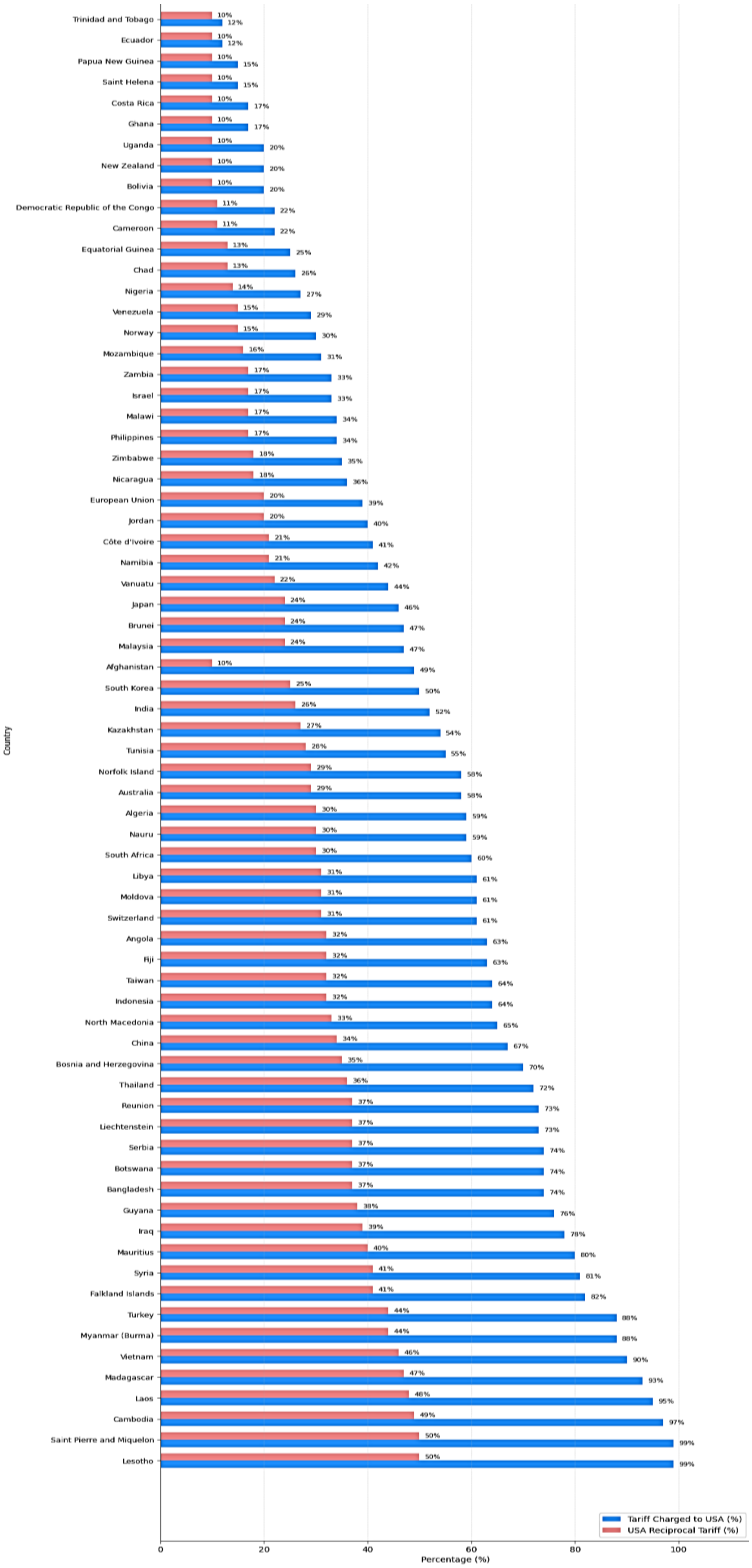

Figure 2: Bar Plot of Full List Comparison of Tariff by Country Under USA Reciprocal Tariff 2025

## 1. Introduction

The United States proposed a reciprocal tariff structure that employed a 'discounted' approach, whereby U.S. tariffs were systematically lower than the tariffs imposed by other countries. This necessitated a rigorous analytical framework to understand the potential systemic impacts such as alterations in trade flows, economic growth, and market dynamics resulting from these tariff adjustments on international trade. This study aimed to decipher the intricate patterns and potential consequences of this tariff regime, particularly in light of its planned implementation in April 2025.

To achieve a comprehensive understanding of these systemic impacts, we quantified the 'discounted reciprocity' pattern inherent in the tariff structure, exploring its strategic implications through the lens of game theory (using linear regression); identified distinct country typologies based on tariff exposure and economic complexity, drawing upon economic complexity theory to understand the policy's differentiated effects (using K-means clustering); and assessed the potential impact on specific sectors, with a focused analysis on the major coffee exporting sector. This sector was particularly relevant as the United States is undeniably a major global importer of coffee, ranking as the world's second-leading importer overall, following the European Union bloc (USDA ERS, 2024). Demonstrating the scale of this demand, the US imported over $8.2 billion worth of coffee in 2023 (Coffee Intelligence, 2025), primarily sourced from South American nations like Colombia ($1.38 billion) and Brazil ($1.35 billion), along with others such as Switzerland (mostly roasted/processed), Canada, Honduras, and Guatemala (USAFacts, 2024). We utilized simulation modeling based on demand elasticity and substitution principles to project potential trade flow shifts and demonstrate how tariff differentials could induce significant substitution effects, potentially altering U.S. import demand in this critical sector.

Throughout, our findings were contextualized within political economy considerations, recognizing the interplay of political and economic forces in shaping trade policy. To visually represent and analyze our findings, we employed a variety of visualization techniques, which included regression plots to illustrate statistical relationships, scatter plots to depict individual data points, cluster maps to identify country groupings, bar charts to compare sector-specific impacts, and case study illustrations to provide in-depth analysis of the coffee export sector. By integrating quantitative techniques like linear regression and K-means clustering with simulation modeling, and grounding our analysis in relevant economic theories such as the gravity model of trade (Tinbergen, 1962) (to understand trade flow patterns), comparative advantage (Ricardo, 1817) (to analyze potential shifts in production and specialization), protectionism (Krugman, 1987) (to contextualize potential motivations behind the tariff policy), and welfare effects (Feenstra, 2015) (to assess the potential overall economic impact), we aimed to provide a comprehensive assessment of the strategic implications and potential systemic impacts of this new tariff structure. The findings of this study offered valuable insights for policymakers considering the design and implementation of reciprocal tariff agreements, enabling them to better understand the potential trade and economic consequences and to make more informed decisions regarding strategic trade policy.

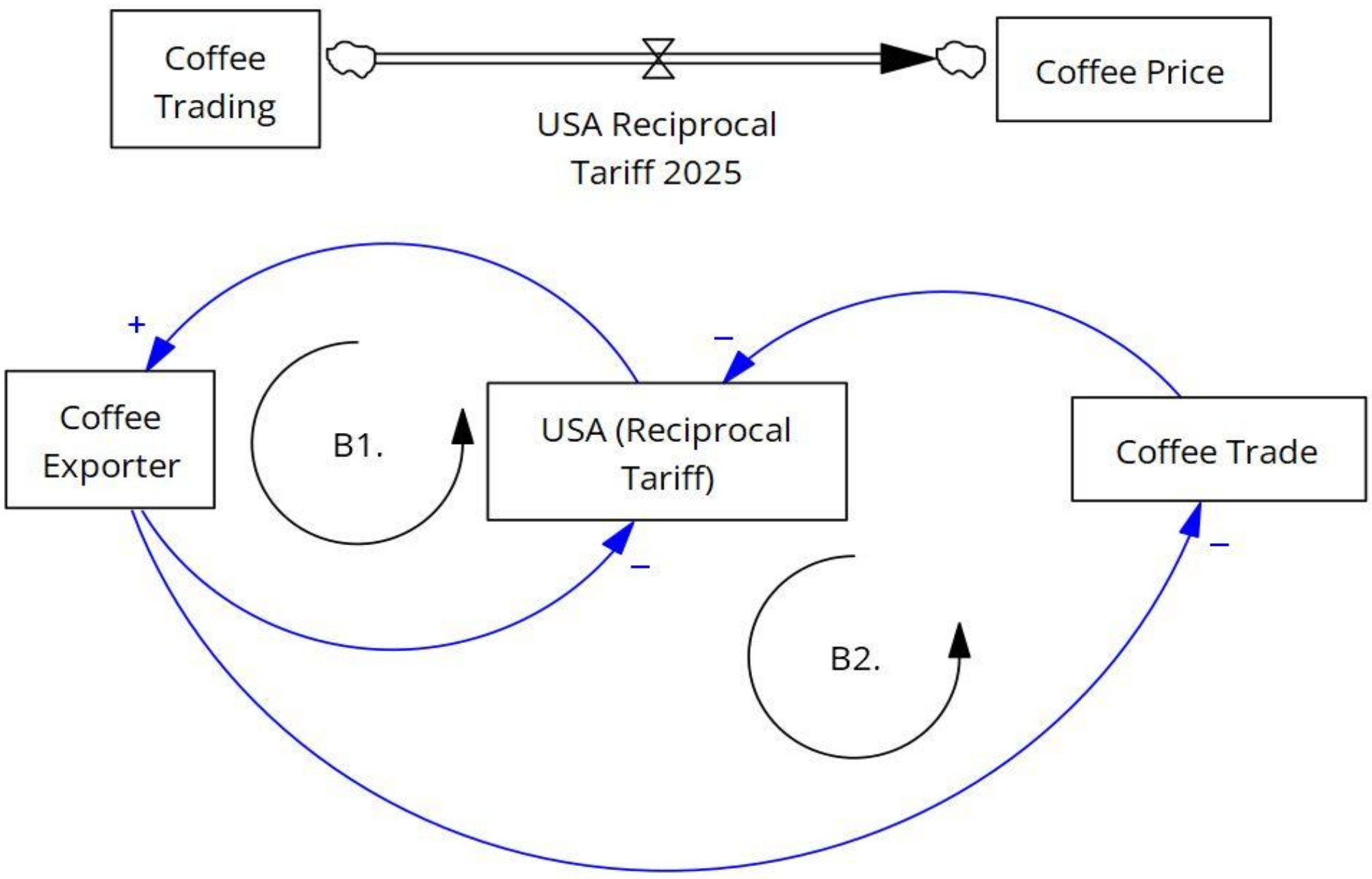

Figure 3: Causal Loop Diagram of Coffee Market Interactions Under USA Reciprocal Tariff 2025

The provided causal loop diagram (CLD) maps the complex interactions expected within the coffee market under the influence of the proposed "USA Reciprocal Tariff 2025." Central to the diagram, the tariff acts as an external pressure that increases the cost of importing coffee, directly impacting trade volume and market prices, thereby disrupting established market dynamics. The CLD further explores the feedback mechanisms triggered by this tariff, detailing two key balancing loops (B1 and B2) that illustrate potential market and political responses. Loop B1, focusing on exporter revenue impact, shows that an increase in the U.S. tariff is expected to decrease the volume of coffee traded; this reduction, in turn, negatively impacts the revenue of coffee exporting countries, and the resulting economic strain on exporters might subsequently generate political or economic pressure advocating for a reduction in the U.S. tariff, completing the balancing feedback. Similarly, loop B2 highlights the tariff's direct negative effect on overall coffee trade volume, illustrating how the broad economic consequences of reduced trade could also mobilize stakeholders (like importers or consumer groups) to lobby for tariff reduction. Overall, the CLD underscores the tariff's anticipated dampening effect on coffee trade. The balancing loops (B1 and B2) illustrate how the system might react, suggesting that negative economic consequences for exporters and disruptions to trade volume could generate counteracting pressures aimed at lowering the tariff barrier. While simplified, this diagram effectively visualizes the interconnectedness of these market variables and potential policy feedback, offering a foundational understanding of the coffee market's dynamic response to the proposed tariff.

## 2. Literature Review

This review established the theoretical framework necessary for a comprehensive analysis of the proposed U.S. reciprocal tariff policy and its complex implications, with a particular focus on the dynamics of the global coffee trade. Given that this study centered on quantifying the degree of 'discounted reciprocity' embedded within the tariff structure and rigorously assessing the resultant shifts in international trade flows, the Gravity Model of Trade (Tinbergen, 1962; Anderson, 1979) assumed a position of central importance. This model provided a robust foundation for not only modelling baseline trade patterns among nations but also for predicting the extent to which the imposition of U.S. tariffs altered the 'frictions' that governed trade relationships with its diverse partners. Its strength lay in its capacity to use factors such as economic size and distance to explain trade volumes, offering a valuable tool to isolate the impact of the tariff policy itself.

Furthermore, to fully comprehend the potential for transformative production and specialization changes that might arise among coffee-exporting nations as a direct consequence of the U.S. tariffs, the classical theories of Comparative Advantage (Ricardo, 1817) were indispensable. These theories enabled a detailed analysis of whether the tariffs served to promote or, conversely, distort the efficient allocation of coffee production across countries based on variations in relative production costs. This was particularly salient in the coffee sector, where diverse growing conditions and production efficiencies already shaped trade patterns.

New Trade Theory (Krugman, 1987; Melitz, 2003; Baldwin & Robert-Nicoud, 2007) significantly enriched the analysis by moving beyond traditional aggregate models to acknowledge the crucial role of firm heterogeneity. Recognizing that coffee firms, both within the U.S. and in exporting countries, exhibited differences in size, productivity, and export capabilities, this theoretical perspective allowed for a more nuanced examination of how the tariffs might influence market competition, economies of scale in coffee processing, and the strategic export decisions of individual firms. This was vital for predicting winners and losers within the coffee industry.

Considering that reciprocal tariffs were inherently strategic instruments, the Political Economy of Trade (Irwin, 1996; Grossman & Helpman) was an essential component of this review. Integrating this perspective allowed for a deeper exploration of the potential motivations underpinning the U.S. policy. These motivations could range from the protection of domestic coffee-related industries to the use of tariffs as leverage in broader trade negotiations. Understanding these political and strategic dimensions was crucial for a holistic evaluation.

Finally, the analysis of the Welfare Effects of Tariffs (Feenstra, 2015; Amiti & Freund, 2010) was of paramount importance. This analysis enabled a thorough evaluation of the broader economic consequences of the tariffs, extending beyond mere trade flows to encompass potential changes in consumer surplus for U.S. coffee consumers, producer surplus for coffee farmers in exporting nations, and the overall efficiency of resource allocation within the global coffee market.

In addition to establishing these theoretical foundations, the review synthesized relevant empirical studies. This synthesis prioritized studies that employed research methodologies closely aligned with those used in this research, including regression analysis, clustering techniques, and simulation modeling. Furthermore, it incorporated studies that specifically analyzed the impact of trade policies on agricultural commodity markets, with a specific and detailed emphasis on the coffee sector. By doing so, the review aimed to provide context for the current study's contributions and clearly identify existing gaps in the literature that this research sought to address.

The methodologies used in this report, such as linear regression, K-Means clustering, and simulation modeling, aligned closely with the theoretical frameworks discussed above. The Gravity Model of Trade was reflected in this report's use of linear regression to quantify the 'discounted reciprocity' pattern in tariffs, consistent with the Gravity Model's approach to understanding trade flow patterns based on economic size and distance. The case study on coffee exporters demonstrated how tariff differentials could induce significant substitution effects, projecting a shift in U.S. import demand away from high-tariff origins like Vietnam towards lower-tariff competitors like Brazil and Colombia. This aligned with the principle of Comparative Advantage, where Brazil might have a relative advantage in coffee production costs under the new tariff regime.

The clustering analysis identified distinct country typologies based on tariff exposure and Economic Complexity Index (ECI), acknowledging firm heterogeneity and the differentiated impacts of tariffs on market competition and strategic export decisions. The report contextualized findings within political economy considerations, recognizing the interplay of political and economic forces in shaping trade policy. This perspective was crucial for understanding the strategic motivations behind the U.S. reciprocal tariff policy. The simulation modeling assessed the potential impact on coffee demand, highlighting the broader economic consequences of the tariffs, including changes in consumer surplus for U.S. coffee consumers and producer surplus for coffee farmers in exporting nations.

## 3. Data Description and Methodology

The proposed U.S. Reciprocal Tariff policy introduces a 'discounted' approach, where the tariffs imposed by the U.S. are consistently lower than those levied by its major trading partners. To quantify this 'discounted reciprocity,' we analyzed the tariff rates applied by the U.S. and its trading partners using data from the World Bank's World Development Indicators (2023). Our analysis focused on the most recent year for which comprehensive data was available. We employed linear regression to model the relationship between U.S. tariffs (dependent variable) and the tariffs imposed by its trading partners (independent variable), expecting to observe a slope less than one to confirm the 'discounted' nature of the reciprocity. Additionally, the Economic Complexity Index (ECI) was incorporated, sourced from the Observatory of Economic Complexity (OEC) (Hausmann et al., 2014), to provide insights into the structural sophistication of each country's export basket. Furthermore, approximate baseline coffee market shares were added, based on estimates from industry reports and trade publications (e.g., International Coffee Organization, 2022). The methodologies employed in this study included linear regression, used to quantify the relationship between tariffs and other economic variables, and K-means clustering, which facilitated the identification of distinct country typologies based on tariff exposure and ECI. A coffee demand shift simulation model was developed, grounded in assumptions regarding price elasticity of demand (PED), tariff pass-through, redistribution effects, and constant demand, to project potential trade flow impacts. Data visualization techniques, including regression plots, scatter plots, cluster maps, bar charts, and case study illustrations, were utilized to effectively present and interpret the findings.

## 4. Analyzing the Reciprocal Tariff Structure

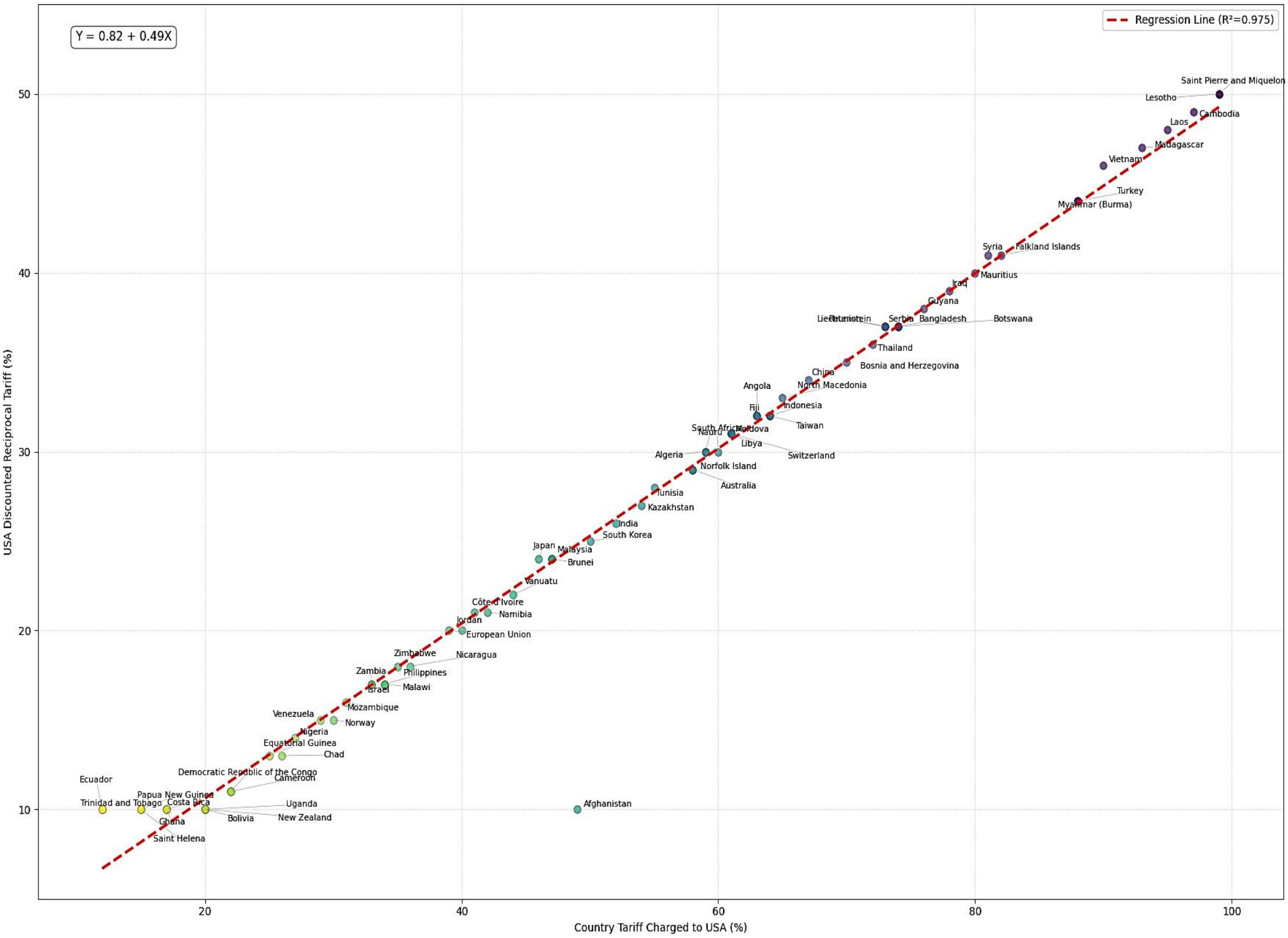

Figure 4: Linear Regression: USA Reciprocal Tariffs 2025

Figure 4 visually represented the results of our regression analysis, plotting the tariff rates of the U.S. against those of its trading partners. This figure served as our testing to prove that the U.S. uses a simple regression model to determine reciprocal tariffs. The regression line illustrated the general trend, indicating a positive relationship where higher tariffs imposed by trading partners were associated with higher, albeit still lower, U.S. tariffs. This visual evidence supported the 'discounted reciprocity' aspect of the policy, quantified as $Y \approx 0.82 + 0.49X$, with a high coefficient of determination ($R^2 \approx 0.975$). This robust correlation indicated a systematic application of the 'discounted' principle in the tariff structure.

To further illustrate this pattern, we examined the line graph depicting tariffs charged versus reciprocal tariffs for top exporters. This graph showcased how the observed linear relationship held across key trading partners, reinforcing the consistent application of the proposed tariff policy. This consistency aligned with strategic trade policy considerations, where predictable and transparent tariff structures could influence trade flows and strategic interactions among nations (Baldwin & Robert-Nicoud, 2007). The high R-squared value suggested that the linear model effectively captured the relationship, indicating that the 'discounted reciprocity' was a dominant feature of the proposed tariff structure.

Additionally, Figure 1 presented a line graph comparing the tariffs charged to the U.S.A. versus U.S.A. discounted reciprocal tariffs across various countries. The blue line, representing "Tariffs Charged to the U.S.A.," generally showed higher tariff percentages compared to the orange line, representing "U.S.A. Discounted Reciprocal Tariffs." This visual representation provided a clear comparison of the average tariff rates of the U.S. and a selection of its major trading partners. This longer-term perspective helped to contextualize the 'discounted reciprocity' proposed under the new policy within historical tariff trends. Afghanistan was the most significant outlier on the graph, located far below the regression line. This indicated that its actual "USA Discounted Reciprocal Tariff (%)" was significantly lower than what the linear model predicted based on its "Country Tariff Charged to USA (%).

## 5. Tariffs vs. Economic Indicators for Top Exporters

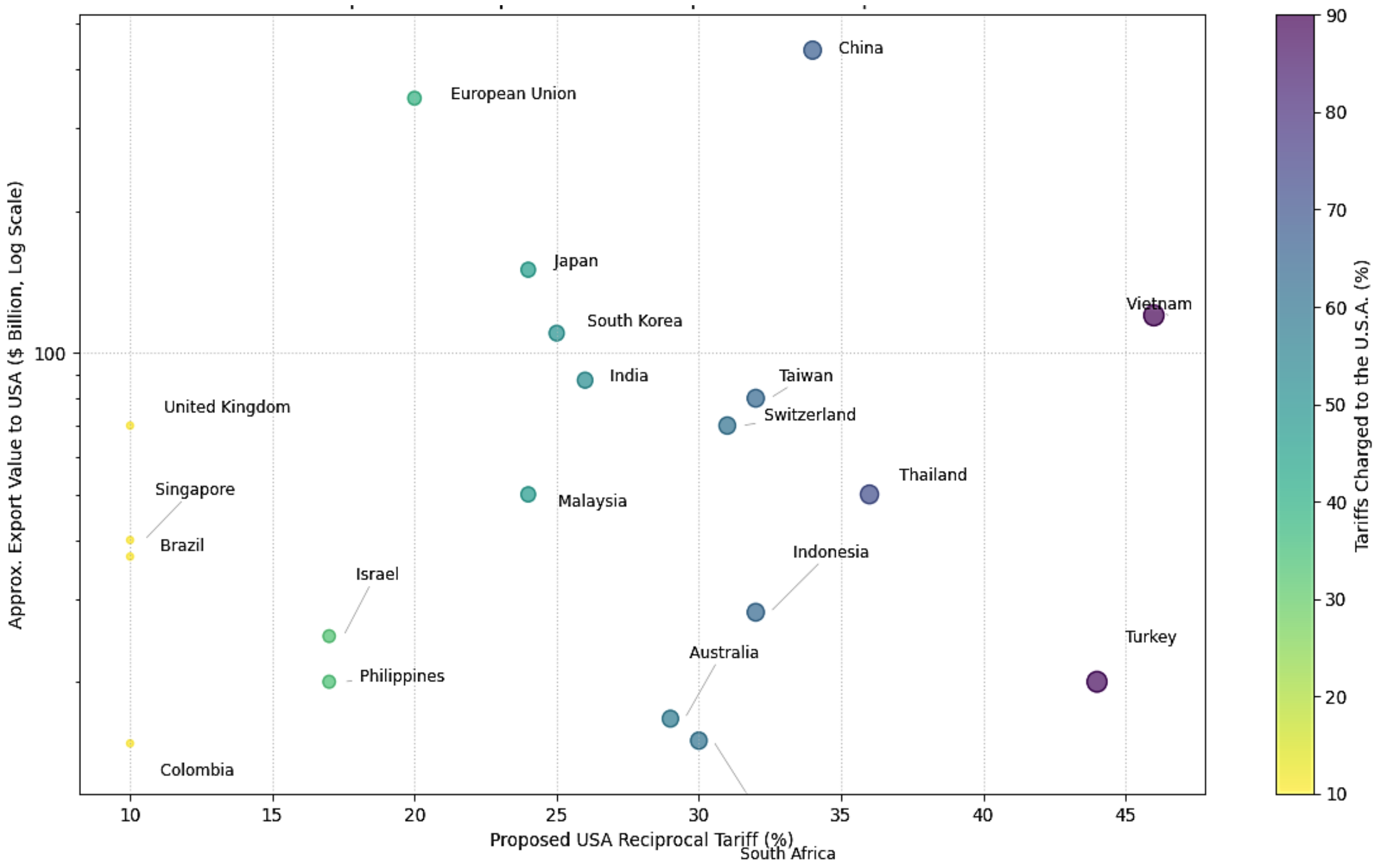

Figure 5: Proposed USA Reciprocal Tariff 2025 vs. Export Value (Top Partners)

Figure 5 further explored the relationship between applied tariffs and economic indicators for a selection of the top U.S. exporting countries. Scatter plots were used to visualize the correlation between tariff rates and metrics such as GDP growth and trade balance, offering insights into whether higher tariffs were associated with specific economic outcomes. The relationship between the proposed reciprocal tariffs and export value for top U.S. trade partners was examined through a scatter plot, where the proposed U.S.A. Reciprocal Tariff (%) was plotted against the approximate export value to the USA ($B), with the latter represented on a logarithmic scale.

The plot revealed a notable dispersion of data points, indicating a lack of strong linear correlation between the proposed tariff levels and the export value. Specifically, countries with relatively low reciprocal tariffs, such as Colombia and Brazil, exhibited significantly lower export values compared to those with higher tariffs, like China. However, this pattern was not consistently observed across all data points. For instance, Vietnam and Turkey, despite having relatively high reciprocal tariffs, did not demonstrate correspondingly high export values, highlighting the complexity of the relationship. Furthermore, the color gradient and size of the points, which represented the tariffs charged to the U.S.A. (%), did not show a direct correspondence with the export value. This suggested that the proposed tariff structure was not solely determined by the sheer volume of trade, as proxied by export value. Rather, it likely reflected a combination of factors, including strategic political economy considerations and industrial policy goals (Irwin, 1996).

The use of a logarithmic scale in representing export value effectively illustrated the wide range of trade scales among the top exporting countries, allowing for the visualization of both minor and major trade partners within the same plot.

## 6. Country Clustering: Tariffs and Economic Complexity

To further understand the differentiated impacts of the proposed tariff policy, we employed a machine learning technique, K-Means clustering, to identify distinct country typologies. This approach allowed us to group countries based on their tariff exposure (average tariff rate imposed by the U.S. and average tariff rate imposed on the U.S.) and their level of economic complexity, as measured by the Economic Complexity Index (ECI). Economic Complexity Index (ECI) analysis was employed to explore the relationship between the structural sophistication of countries' export baskets and the proposed U.S.A. Reciprocal Tariffs.

The cluster plot, depicting USA Reciprocal Tariff (%) vs. ECI, revealed four distinct country clusters, each defined by convex hulls and characterized by unique attributes as detailed in the legend. Figures 6 and 7 present cluster maps visualizing the country groupings identified through this analysis. These maps illustrate the geographical distribution of countries within each cluster, providing a visual representation of how tariff exposure and economic complexity correlate geographically.

Group 1, represented by the yellow convex hull, exhibits 'High ECI / Low Tariff' and primarily comprises countries with high economic complexity and low reciprocal tariffs. Group 2, shown within the purple convex hull, is characterized by 'High ECI / High Tariff,' encompassing countries with high economic complexity and high reciprocal tariffs. Group 3, enclosed by the green convex hull, represents 'Low ECI / High Tariff,' including nations with low economic complexity and high reciprocal tariffs. Finally, Group 4, within the blue convex hull, signifies 'Low ECI / Low Tariff,' featuring countries with low economic complexity and low reciprocal tariffs.

These distinct clusters, delineated by convex hulls, effectively illustrate the data distribution and the relationships between reciprocal tariff levels and ECI. The clustering of countries with high ECI suggests that these economies, as predicted by complexity theory (Hausmann & Hidalgo, 2009), are better able to adapt to tariff shocks due to their diversified export baskets. This clustering analysis, therefore, provides a valuable framework for understanding the potential distributional effects of the proposed tariff regime across countries with varying economic structures, aligning with theories of economic development and trade that emphasize the role of ECI in adapting to trade policy changes (Hidalgo & Hausmann, 2009).

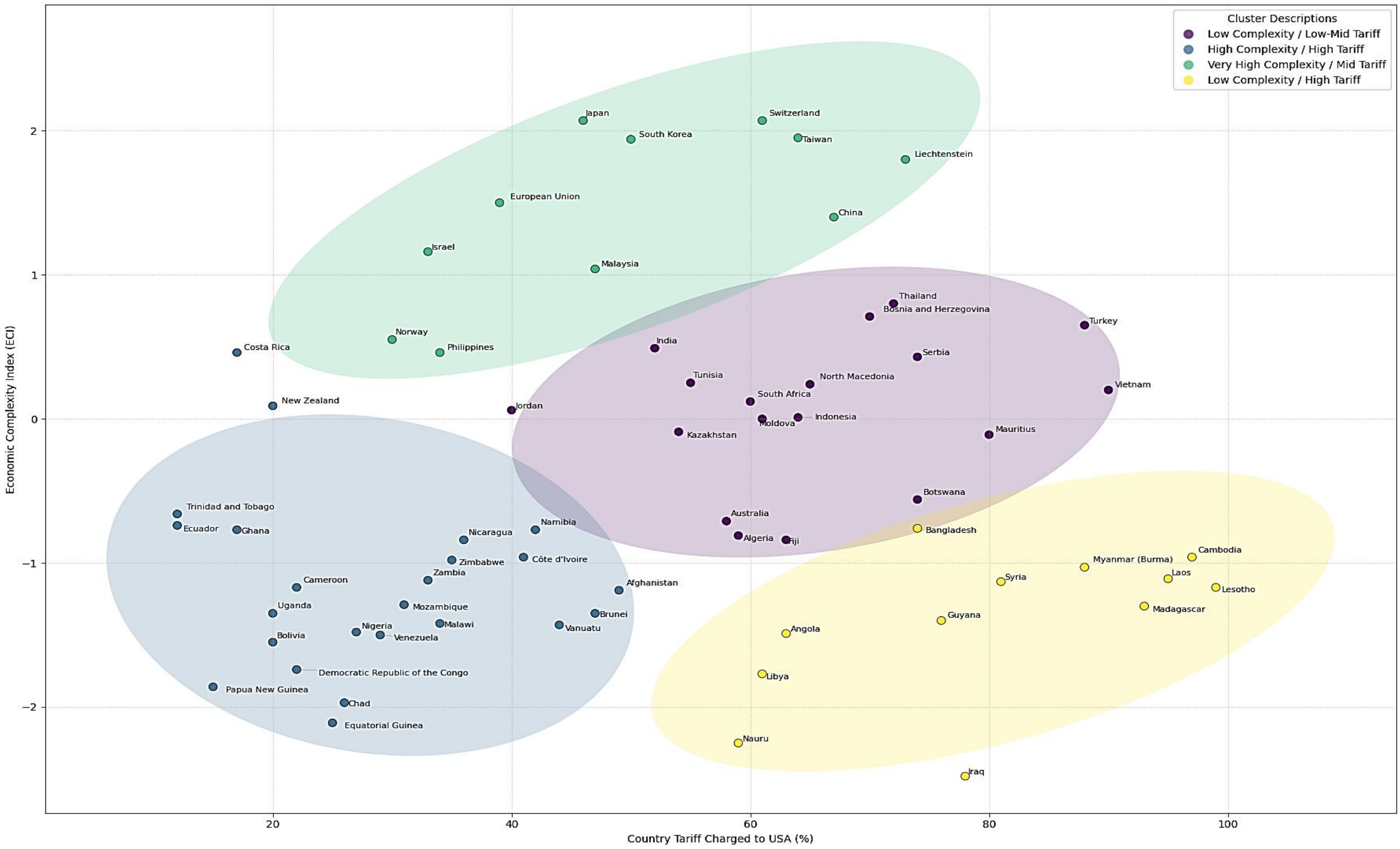

Figure 6: Country Clusters with Descriptive Names (Tariff vs ECI, k=4)

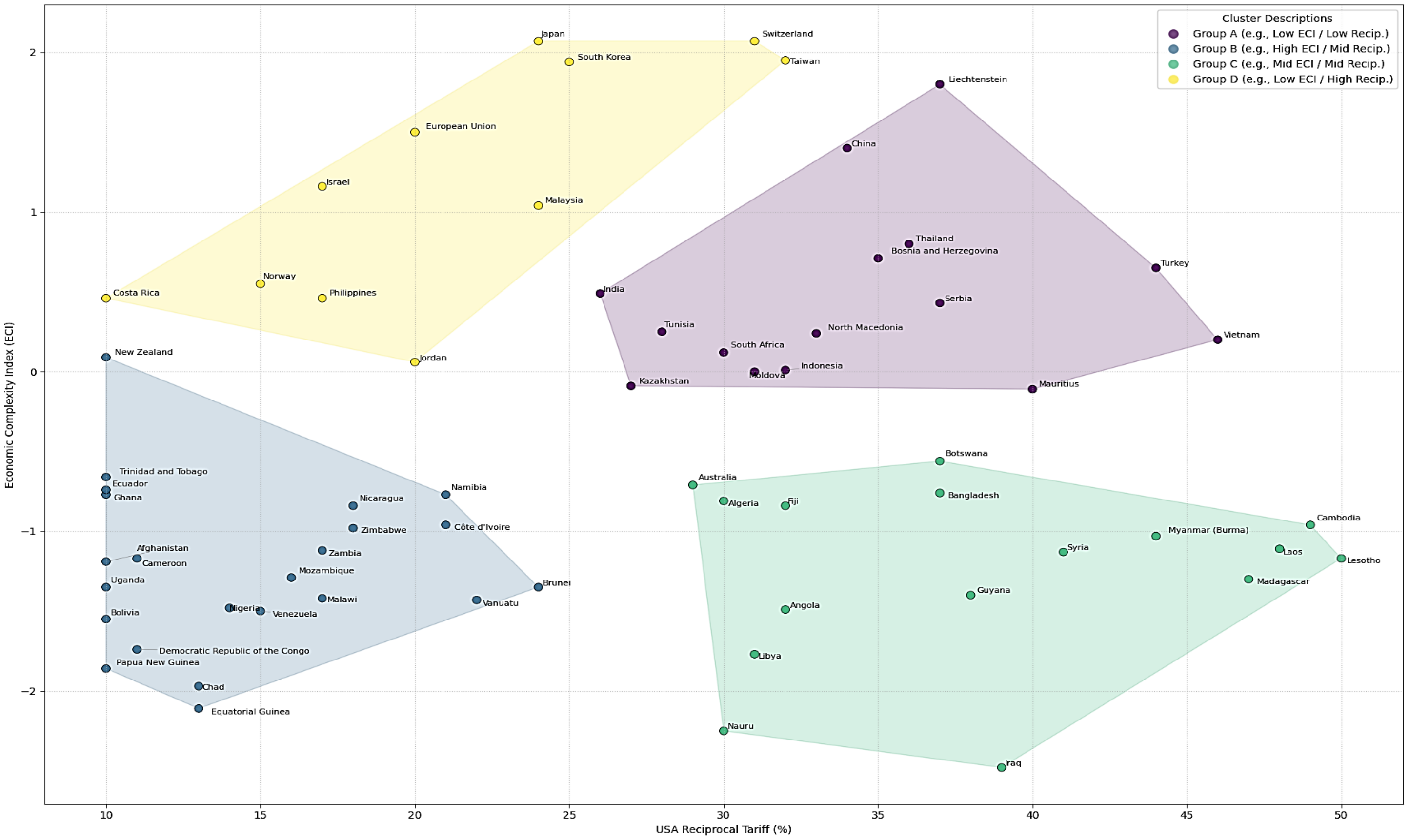

Figure 7: Country Clusters with Convex Hulls (Reciprocal Tariff vs ECI, k=4)

## 7. Case Study: Impact Assessment for Coffee Exporters

To assess the potential impact on specific sectors, we focused on the major coffee exporting sector. Figures 8, 9, 10, and 11 illustrate the findings of our simulation modeling. Figure 8 presents a comparative view of tariff situations for major coffee exporting countries. The scatter plot shows the relationship between the tariffs charged to the USA by various countries and the proposed US reciprocal tariffs. The plot includes a dashed line indicating equal tariffs (y = x) and a solid blue line indicating a discounted scenario (y = 0.5x). Countries such as Vietnam, Indonesia, India, Nicaragua, Ethiopia, Guatemala, Costa Rica, Uganda, Peru, Honduras, and Colombia are plotted, with Vietnam showing the highest proposed US reciprocal tariff at approximately 40%. This visual representation highlights the significant tariff differentials faced by exporters like Vietnam compared to others, setting the stage for analyzing projected trade flow shifts driven by demand elasticity and substitution effects.

Figure 9 illustrates the impact of tariffs on the demand for Vietnamese coffee in the U.S. market, demonstrating how different elasticity scenarios affect price and demand. The plot shows three scenarios: the initial state with price (P) and quantity (Q) both set to 100, the scenario after the tariff with inelastic demand (PED = 0.5) showing a 34.5% price increase due to a 4% tariff, with a 17.2% change in price and a 51.8% decrease in quantity demanded, and the scenario after the tariff with elastic demand (PED = 1.5) also showing a 34.5% price increase due to a 4% tariff, but with a more significant impact on quantity demanded due to higher price elasticity of demand. This figure vividly demonstrates the price and demand impact of Vietnam's tariff under different elasticity scenarios, highlighting the importance of demand elasticity in understanding the effects of tariff changes on market dynamics.

Figure 10 illustrates the cost competitiveness of different coffee origins under the proposed tariff regime. The graph, titled "Relative Cost of Coffee Origins vs. Vietnam (Flipped Axes)," compares the illustrative cost index of various coffee origins to that of Vietnam, assuming a 75% tariff pass-through. The x-axis represents potential substitute coffee origins, including Peru, India, Costa Rica, Colombia, Brazil, Ethiopia, and Honduras, while the y-axis represents the illustrative cost index, with a base of 100. The dashed red line at the top indicates the Vietnam cost index of 134.5. Countries with lower y-values are considered cheaper alternatives to Vietnam under the assumed tariffs. For example, Peru, India, and Costa Rica have lower cost indices compared to Vietnam, making them more cost-competitive under the proposed tariff regime. This figure highlights how tariff differentials influence the cost competitiveness of various coffee origins, providing insights into potential shifts in U.S. import demand based on relative costs.

Finally, Figure 11 projects the potential shift in U.S. import demand for coffee, highlighting the 'discounted' nature of the proposed U.S. reciprocal tariffs and the significant tariff differentials faced by exporters like Vietnam compared to others. The bar chart compares the estimated market share of major coffee exporting countries before and after the implementation of the proposed tariffs. The x-axis represents major coffee exporting countries, while the y-axis represents the share of U.S. coffee imports in percentage terms. The chart clearly illustrates the projected changes in U.S. import market shares for these countries. Notably, it shows a significant decrease in Vietnam's market share, accompanied by a substantial increase in the market shares of Brazil and Colombia. This shift indicates a projected movement of U.S. coffee import demand away from high-tariff origins, such as Vietnam, towards low-tariff origins, like Brazil and Colombia. This projection underscores the potential impact of the proposed tariffs on trade flows, driven by demand elasticity and substitution effects. The simulation, based on assumptions of PED=-1.5, 75% price pass-through, and proportional redistribution of lost shares, aligns with economic theories of trade diversion, where tariffs can lead to the substitution of imports from efficient, high-tariff countries to less efficient, low-tariff countries (Viner, 1950). The simulation of demand shifts, based on elasticity and substitution, provides a quantitative assessment of the potential sectoral impacts of the proposed tariff regime.

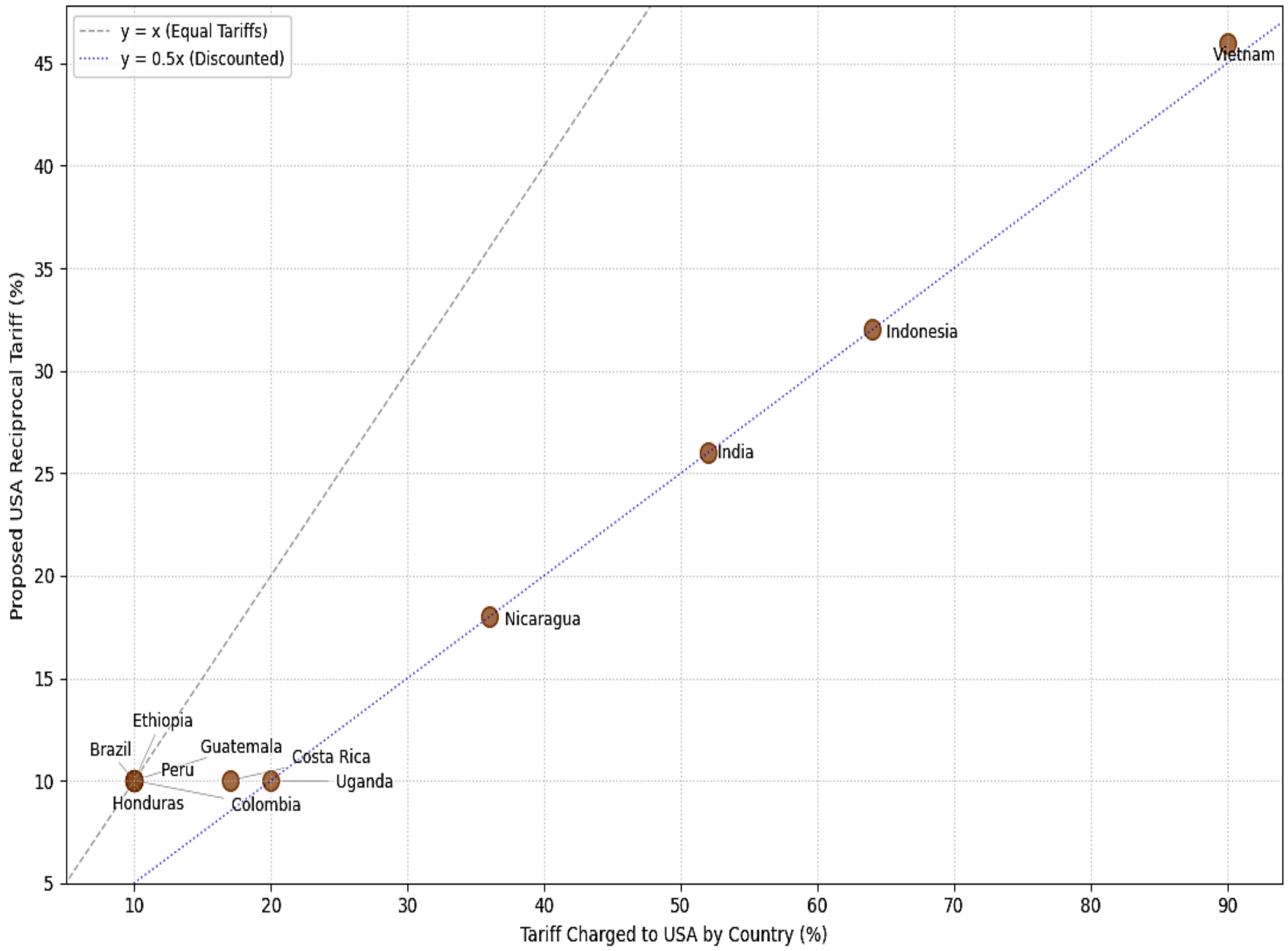

Figure 8: Potential USA Reciprocal Tariffs 2025 for Major Coffee Exporters

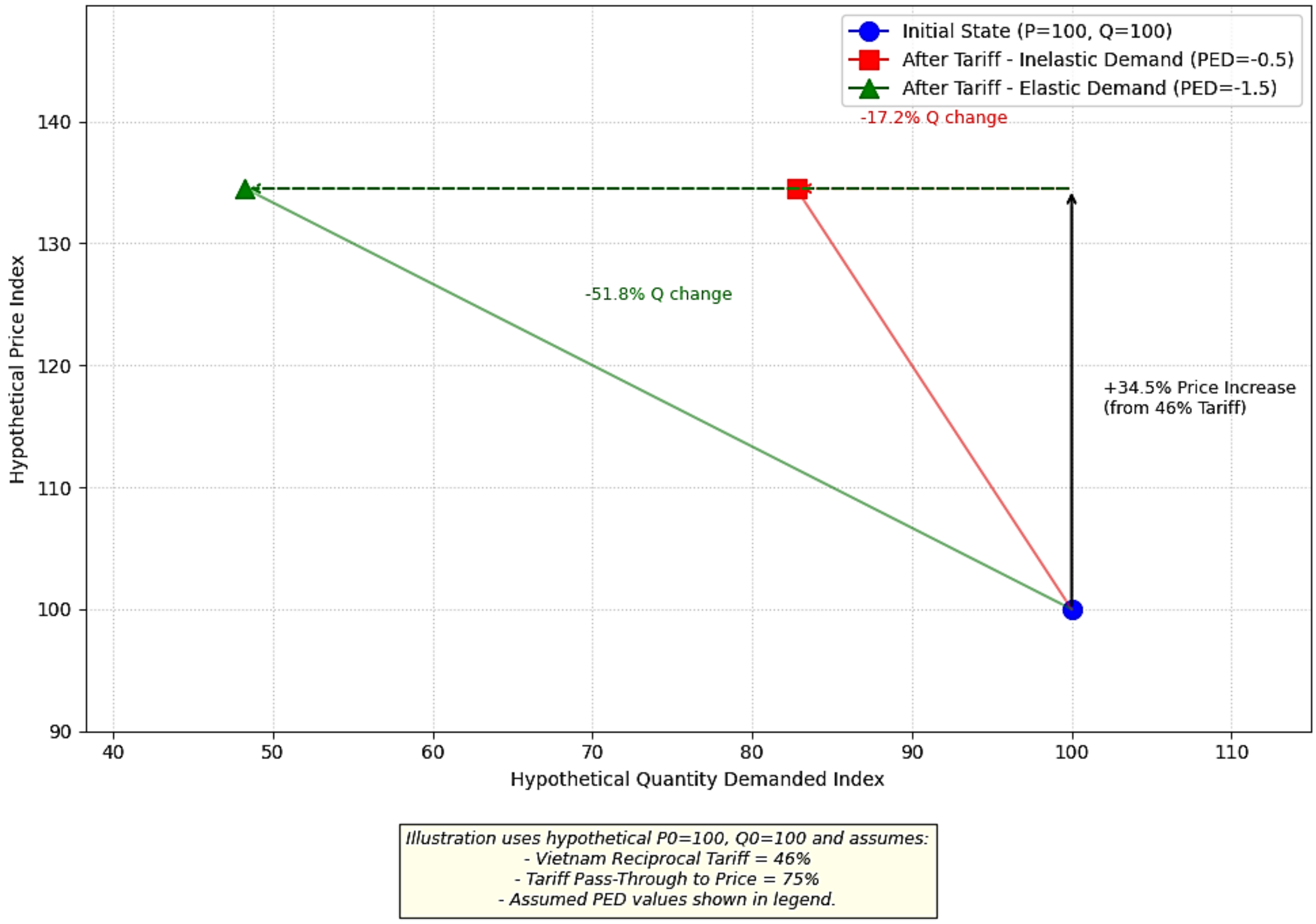

Figure 9: Impact of USA Reciprocal Tariffs 2025 on Demand: Vietnam Coffee Industry Example

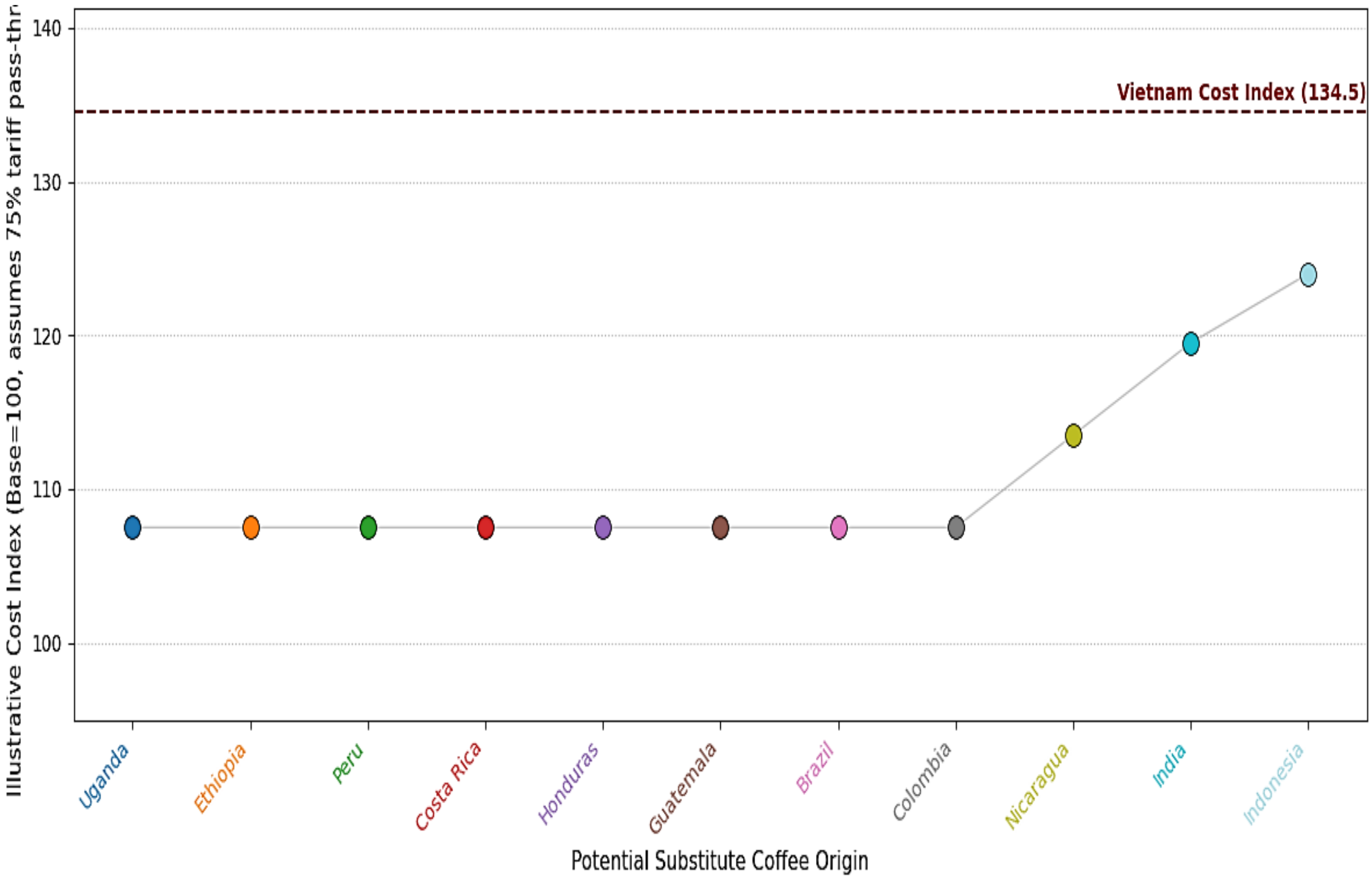

Figure 10: Hypothetical Relative Cost of Coffee Origins vs. Vietnam Post USA Reciprocal Tariffs 2025

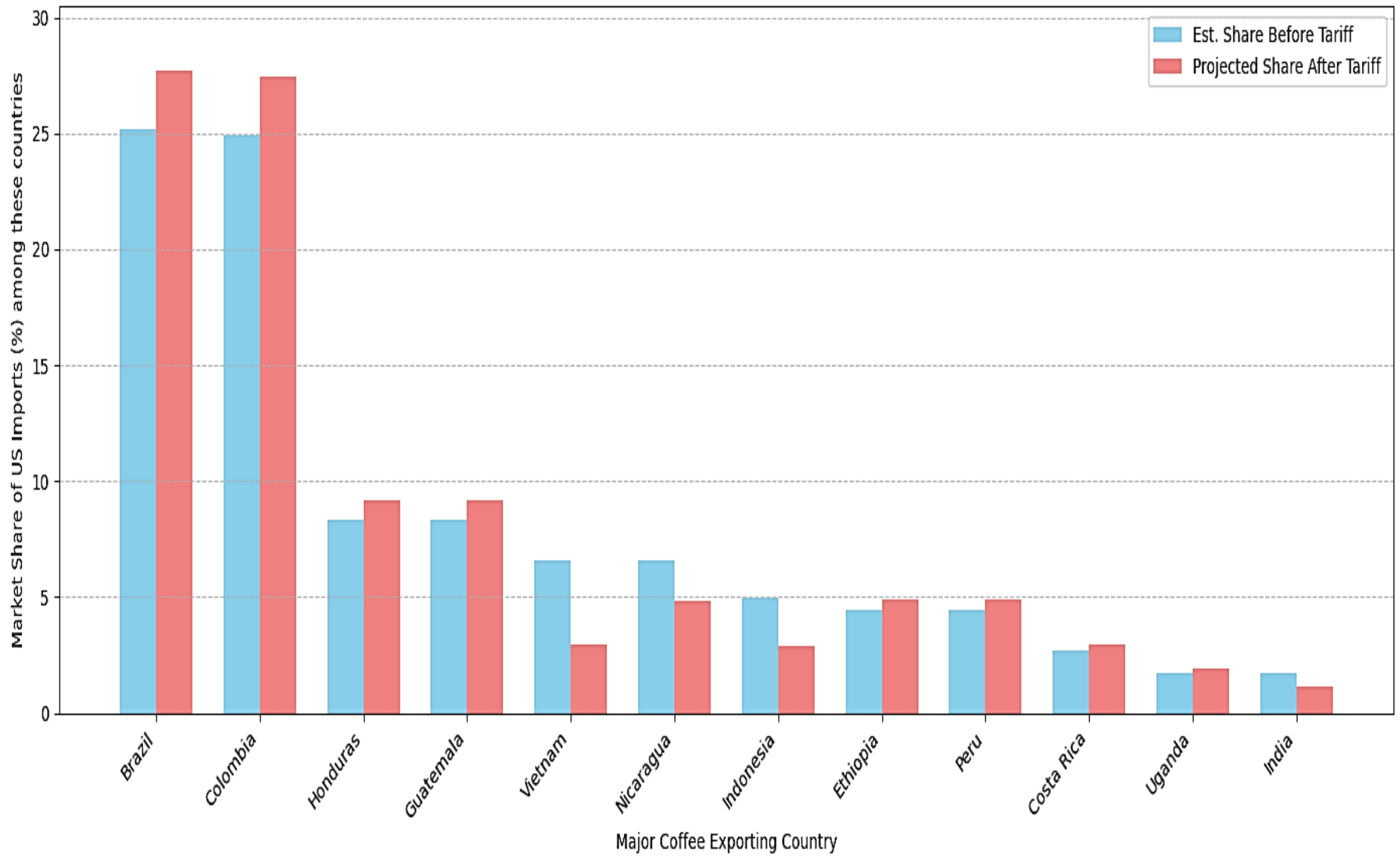

Figure 11: Illustrative Projection of Coffee Demand Shift Due to Proposed USA Reciprocal Tariffs 2025

## 8. Discussion

The analysis reveals a clear 'discounted reciprocity' pattern in the proposed U.S. tariff structure, strongly supported by linear regression results (Baldwin & Robert-Nicoud, 2007). This pattern indicates that the U.S. systematically imposes lower tariffs compared to those levied by its trading partners, aligning with strategic trade policy considerations. While reciprocal tariffs do not exhibit a direct correlation with export value, suggesting influences beyond trade volume (Irwin, 1996), the clustering analysis demonstrates a significant relationship with the Economic Complexity Index (ECI). This relationship indicates that countries with higher economic complexity tend to experience differentiated impacts from the tariff policy, reflecting their varied levels of economic diversification (Hidalgo & Hausmann, 2009).

The case study on the coffee sector highlights the potential for substantial trade flow shifts due to tariff-induced substitution effects. The simulation modeling projects a significant shift in U.S. import demand away from high-tariff origins like Vietnam towards lower-tariff competitors such as Brazil and Colombia. This outcome is well-explained by theories of trade diversion (Viner, 1950), where tariffs lead to the substitution of imports from efficient, high-tariff countries to less efficient, low-tariff countries. The shift in coffee imports from Vietnam to Brazil aligns with the principle of Comparative Advantage (Ricardo, 1817), as Brazil may have a relative advantage in coffee production costs under the new tariff regime.

The application of Game Theory, Political Economy, and Complexity theories proves suitable for interpreting the strategic interactions, political influences, and structural economic factors underpinning the observed tariff patterns and projected impacts. Game Theory helps explain the strategic behaviour of countries in response to the tariff policy, while Political Economy considerations highlight the interplay of political and economic forces in shaping trade policy. Complexity theory underscores the importance of economic diversification in adapting to tariff shocks.

However, it is important to acknowledge the limitations of this study. The reliance on hypothetical tariff data and approximations in external data sources introduces potential uncertainties. Additionally, the inherent assumptions within the simulation models, such as price elasticity of demand and tariff pass-through rates, may affect the accuracy of the projections. Future research should aim to validate these findings with real-world data and explore the dynamic impacts of these tariffs over time.

The imposition of a 46% reciprocal tariff on Vietnamese coffee imports is likely to raise U.S. consumer prices for coffee. This significant tariff increase will directly impact the cost of importing coffee from Vietnam, one of the major coffee exporters to the U.S. Importers and retailers are expected to pass on these increased costs to consumers, resulting in higher prices for coffee on the shelves. The tariff-induced price increase may also lead to a shift in demand towards coffee from countries with lower tariffs, such as Brazil and Colombia. However, this adjustment period could still see higher prices as the market rebalances. Furthermore, the sudden imposition of high tariffs can disrupt existing supply chains, leading to short-term shortages and price volatility. Overall, while the market may eventually adjust, U.S. consumers can expect to see higher coffee prices in the near term due to the reciprocal tariffs on Vietnamese coffee.

In summary, the proposed reciprocal tariff policy is characterized by a consistent 'discounted reciprocity' pattern, with significant implications for trade flows and consumer prices. The findings underscore the importance of considering both Economic Complexity and potential trade diversion effects when implementing such tariff structures. Future research could explore the broader geopolitical implications of the observed strategic trade patterns and the long-term impacts on global trade dynamics.

## 9. Conclusion

The analyses reveal that the proposed reciprocal tariff policy is characterized by a consistent 'discounted reciprocity' pattern, demonstrating a systematic reduction in U.S. tariffs relative to those imposed on the U.S. This pattern is strongly supported by linear regression results, indicating a strategic approach to tariff imposition that aligns with broader trade policy objectives. The clustering analysis further highlights the differentiated impacts of the tariff policy across countries with varying levels of economic complexity, suggesting that more economically complex countries are better positioned to adapt to these changes.

A primary consequence highlighted by the coffee case study is the potential for significant trade flow redirection. The simulation modeling projects a shift in U.S. import demand away from high-tariff origins like Vietnam towards lower-tariff competitors such as Brazil and Colombia. This shift is driven by demand elasticity and substitution effects, where tariffs lead to the substitution of imports from efficient, high-tariff countries to less efficient, low-tariff countries. This outcome aligns with economic theories of trade diversion and comparative advantage, underscoring the strategic implications of the proposed tariff policy.

The imposition of a 46% reciprocal tariff on Vietnamese coffee imports is likely to raise U.S. consumer prices for coffee. This significant tariff increase will directly impact the cost of importing coffee from Vietnam, one of the major coffee exporters to the U.S. Importers and retailers are expected to pass on these increased costs to consumers, resulting in higher prices for coffee on the shelves. The tariff-induced price increase may also lead to a shift in demand towards coffee from countries with lower tariffs, such as Brazil and Colombia. However, this adjustment period could still see higher prices as the market rebalances. Furthermore, the sudden imposition of high tariffs can disrupt existing supply chains, leading to short-term shortages and price volatility. Overall, while the market may eventually adjust, U.S. consumers can expect to see higher coffee prices in the near term due to the reciprocal tariffs on Vietnamese coffee.

Policy implications stemming from this research suggest the importance of considering both Economic Complexity and potential trade diversion effects when implementing such tariff structures. Policymakers should be aware of the broader economic consequences, including potential increases in consumer prices and disruptions to supply chains. Future research could explore the dynamic impacts of these tariffs over time, as well as the broader geopolitical implications of the observed strategic trade patterns. Additionally, further studies could validate these findings with real-world data and examine the long-term effects on global trade dynamics and economic development.

By integrating quantitative techniques with economic theory, this study provides a robust framework for assessing the potential systemic impacts of strategic trade policies. The findings offer valuable insights for policymakers, enabling them to make more informed decisions regarding the design and implementation of reciprocal tariff agreements.

In light of these findings, it is crucial to consider the potential social ramifications of rising coffee prices. While it is unlikely that we will witness a modern-day equivalent of the Boston Tea Party, the historical precedent serves as a reminder of the profound impact that commodity prices can have on social stability. Ensuring that such economic policies are carefully crafted and implemented can help mitigate the risk of social unrest and maintain consumer confidence.